\documentclass[prl, reprint]{revtex4-1}

\usepackage{graphicx}
\usepackage[latin1]{inputenc}
\usepackage{amsmath}
\usepackage{amsfonts}
\usepackage{amssymb}
\usepackage{color}

\begin{document}

\newcommand{\eu}{$^{151}$Eu$^{3+}$:Y$_2$SiO$_5$}
\newcommand{\euo}{$^{153}$Eu$^{3+}$:Y$_2$SiO$_5$}
\newcommand{\tranz}{$\pm|3/2\rangle_g \rightarrow \pm|5/2\rangle_e$ }
\newcommand{\trann}{$\pm|5/2\rangle_g \rightarrow \pm|3/2\rangle_e$ }
\newcommand{\trans}{$\pm|1/2\rangle_g \rightarrow \pm|5/2\rangle_e$ }
 \newcommand{\transition}{$^7$F$_0 \rightarrow ^5$D$_0$ }

\newcommand{\nweak}{2.5 \pm 0.6}
\newcommand{\nweakB}{11.2 \pm 0.6}
\newcommand{\etas}{$(3.8 \pm 1.5 )\cdot 10^{-3} $}
\newcommand{ \pnoise}{$(7.1 \pm 2.3)\cdot 10^{-3} $}

\title{Single-photon-level optical storage in a solid-state spin-wave memory}

\author{N. Timoney}
\author{I. Usmani}
\author{P. Jobez}
\author{M. Afzelius}
\email{mikael.afzelius@unige.ch}
\author{N. Gisin}

\affiliation{Group of Applied Physics, University of Geneva, CH-1211
  Geneva 4, Switzerland}

\begin{abstract}
A long-lived quantum memory is a firm requirement for implementing a quantum repeater scheme. Recent progress in solid-state rare-earth-ion-doped systems justifies their status as very strong candidates for such systems. Nonetheless an optical memory based on spin-wave storage at the single-photon-level has not been shown in such a system to date, which is crucial for achieving the long storage times required for quantum repeaters. In this letter we show that it is possible to execute a complete atomic frequency comb (AFC) scheme, including spin-wave storage, with weak coherent pulses of \mbox{$\bar{n} = \nweak$} photons per pulse. We discuss in detail the experimental steps required to obtain this result and demonstrate the coherence of a stored time-bin pulse. We show a noise level of \mbox{\pnoise} photons per mode during storage, this relatively low-noise level paves the way for future quantum optics experiments using spin-waves in rare-earth-doped crystals.
\end{abstract}


\maketitle

Quantum communication if rigorously executed provides us with a provably secure method of communication \cite{Gisin2002}. However, inherently lossy channels limit the distance over which the communication can be performed, which today is roughly 250 km \cite{Stucki2009,Wang2012}. A quantum repeater which can in principle allow quantum communication over longer distances \cite{Briegel1998,Simon2007,Sangouard2011}, provided that the required quantum memories are developed. Prime candidates for quantum memories are atomic systems, which are capable of maintaining the coherence of stored excitations for long times. Atomic systems that are currently investigated range from individual quantum systems\cite{Specht2011, Piro2011}, laser-cooled atomic gases \cite{Dudin2010,Bao2012}, room-temperature atomic vapours \cite{Julsgaard2004,Reim2011,Hosseini2011}, to rare-earth-ion-doped crystals \cite{Clausen2011,Saglamyurek2011}.

Crystals doped with rare-earth-ion impurities have attractive coherence properties when cooled $<4$K, in particular hyperfine states can have coherence times which can approach seconds \cite{Longdell2005}. This has provided a strong motivation for developing quantum memories using such systems. Following the first storage experiment at the single-photon level \cite{Riedmatten2008}, a succession of experiments demonstrated storage of single photons \cite{Saglamyurek2012,Clausen2012}, generation of light-matter \cite{Clausen2011, Saglamyurek2011} and matter-matter entanglement using crystals \cite{Usmani2012}. The quantum memory performances have also been strongly developed, particularly in terms of storage efficiency \cite{Hedges2010,Sabooni2013}, multimode capacity \cite{Usmani2010,Bonarota2011a} and polarization qubit storage \cite{Clausen2012,Zhou2012,Gundogan2012}.

These experiments were performed for short storage times (in the 10 ns to few $\mu$s regime)  using an optical coherence, rather than exploiting long spin coherence times. Spin storage experiments require strong optical control fields to convert the initial optical coherence to a spin coherence. Photon noise is induced by such an operation, which has been nonetheless shown to work for alkali atomic systems \cite{Specht2011, Piro2011,Dudin2010,Bao2012,Reim2011,Hosseini2011}. In rare-earth-ion-doped solids the task is complicated since there is less spectral separation between the weak signal field and the optical control field (roughly 100 times less). Scattering from the control field is thus more likely, as it propagates through a dense solid-state crystal.

Two quantum memory schemes were specifically proposed for solid-state ensembles; the controlled and reversible inhomogeneous broadening (CRIB) memory (see \cite{Tittel2010b} and references therein) and the atomic frequency comb (AFC) memory \cite{Afzelius2009a}. The AFC has a particularly high multimode capacity, which is the ability to store trains of single photon pulses \cite{Afzelius2009a,Nunn2008}. This is crucial for speeding-up quantum repeater protocols \cite{Simon2007}. The AFC scheme is based on an echo induced by a regular spectral grating of periodicity $\Delta$, in the absorption profile of an atomic ensemble. An AFC echo is emitted a time defined by $1/\Delta$, unless the optical coherence is transferred (written) to a spin coherence before the time $1/\Delta$ has elapsed. Reversing the transfer retrieves an optical pulse (referred to as an AFC spin-wave echo). AFC memories which only use the optical coherence are delay lines unless combined with spin-wave storage  \cite{Afzelius2009a}, which allow for on-demand read out and significantly longer storage times. Only a few AFC spin-wave storage experiments have been reported, all involving storage of bright classical pulses \cite{Afzelius2010,Timoney2012,Gundogan2013}.

Here we demonstrate storage of an optical pulse containing a few photons on average, using an AFC memory combined with spin-wave storage in a europium doped Y$_2$SiO$_5$ crystal. We apply a strategy of filtering in space, time and frequency in order to reduce unwanted emission from the crystal at the moment the weak pulse is recovered from the crystal. To quantify the degree of noise we measure the unconditional noise floor \cite{Reim2011},which is the probability for the memory to produce a noise photon when the memory is read. We report that the unconditional noise floor can be reduced to \pnoise $~$ by our filtering strategy, which is low enough to allow for a range of quantum information schemes that require manipulation of spin coherence. Using the ability of the AFC memory to store multiple time bins, we also store and analyse a time-bin pulse with higher photon numbers, showing the high coherence of our quantum memory.

\begin{figure}
\centering
{\includegraphics[width = 0.5\textwidth]{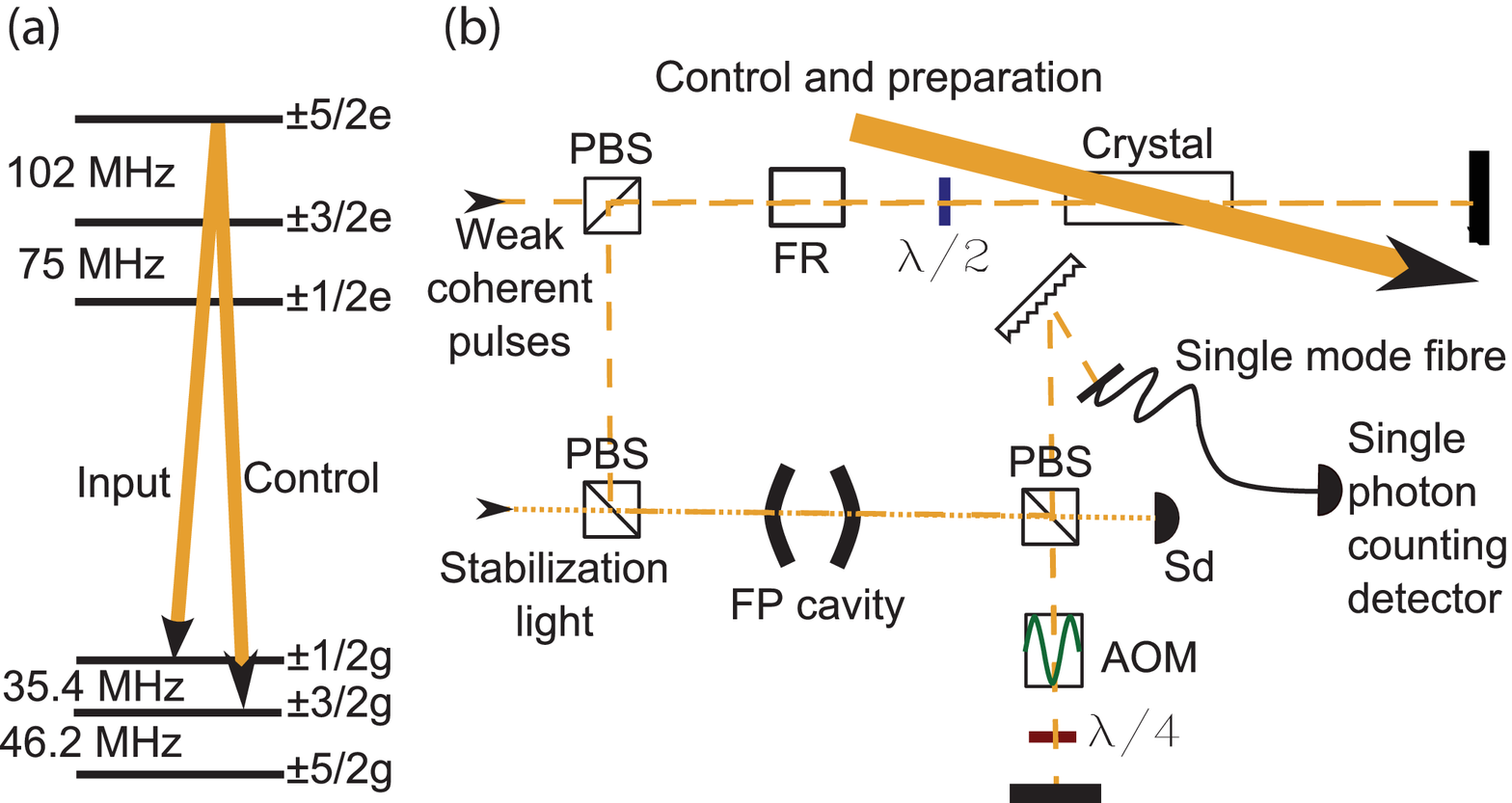}}
\caption{(a)The atomic level scheme of the optical transition \transition  in \eu. (b) A schematic of the experimental setup around the memory, the rest of the experiment has been suppressed for simplicity. The control and preparation beam is in single pass (wide labelled arrow). The input mode (thin dashed line) is in double pass, with the help of a Faraday Rotator (FR) and a polarizing beamsplitter (PBS). On return from the crystal the input mode passes through a Fabry-Perot(FP) cavity (bandwidth of 7.5 MHz). A classical detector (Sd) and 10 $\mu$W of horizontally polarized light (thin dotted line) is used to actively and intermittently stabilize the cavity to the frequency of the input mode. An accousto optical modulator (AOM) in double pass acts as a detector gate.}
\label{fig:expsetup}
\end{figure}

Europium is a promising candidate for quantum memories due to its fine coherence properties at T $<$ 6K \cite{Equall1994,Koenz2003,Alexander2007}, which ultimately could lead to an extremely long-lived \cite{Longdell2006} and multimode memory \cite{Afzelius2009a}. In this work we use the optical \transition transition at 580 nm. The crystal is isotopically pure \eu (100ppm). At a temperature of around 3 K we measure an overall absorption coefficient of $\alpha$ = 1.5 cm$^{-1}$ and an optical inhomogeneous linewidth of 500 MHz. The relevant energy diagram is shown in Fig. \ref{fig:expsetup}a. Our input and control fields excite two optical-hyperfine transitions separated by 35.4 MHz.

The schematic of the experimental set up (Fig. \ref{fig:expsetup}b) shows only the optics around the cryostat containing the \eu $~$ crystal of length $L$=1 cm. The storage mode crosses the control and preparation mode through the crystal. Given the measured angular separation of the beams before the cryostat we estimate a spatial mode overlap of 95 $\%$. A double-pass configuration was implemented on the storage mode to increase the optical depth \cite{Usmani2010}, while the control mode was in single pass. The laser and the acousto-optic modulators (AOMs) used for spectral control are not shown in Fig. \ref{fig:expsetup}b. The laser at 580 nm is a commercially available system based on an amplified diode laser at 1160 nm and a frequency doubling stage.
Before the cryostat the intense control pulses had peak powers of up to 300 mW. The diode laser is stabilized to  have a spectral linewidth of approximately 30 kHz.

The AFC comb structures are created with frequency selective optical pumping techniques, which are now well-established techniques for spectral shaping of inhomogeneously broadened transitions, see for instance \cite{Afzelius2010,Lauritzen2012}. A particular feature of our preparation sequence is that we first pump all ions into the $\pm|1/2\rangle_g$ state, and then create the comb-structure by removing atoms from this state. This has the benefit of reducing the effect of the inhomogeneous spin linewidth, which could otherwise limit the minimum tooth width in the comb structure \cite{Timoney2012}. The maximum optical depth we can achieve on the input transition is $\alpha L$ = 2.4, in double-pass configuration.

We first characterize our memory using bright input pulses of many photons and detecting the pulses with a linear photodiode. We observe AFC echo efficiencies of more than 5$\%$ for \mbox{1/$\Delta$ = 6 $\mu$s},  and {AFC spin-wave} echo efficiencies of 1$\%$ for spin-wave storage time $T_S$ of 18 $\mu$s. The reduction in efficiency is mostly due to imperfect control pulses. We estimate the transfer efficiency per control pulse to be 0.49. By measuring the decay of the spin-wave echo as a function of $T_S$, we estimate the inhomogeneous spin linewidth to be 8 kHz. This measurement will be further detailed in a future publication. The 8 kHz linewidth is surprisingly low, a factor of 8 less than for the \euo (100 ppm) sample we previously used \cite{Timoney2012}. This results in a spin-wave memory time of about 50 $\mu$s, defined by the point where the efficiency is reduced to $\exp(-1)$, the longest so far obtained in an AFC memory. By applying spin refocussing techniques we can expect to increase it further, up to the spin coherence time of 15 ms \cite{Alexander2007}.

\begin{figure*}
    \centering
    \includegraphics[width = 1\textwidth]{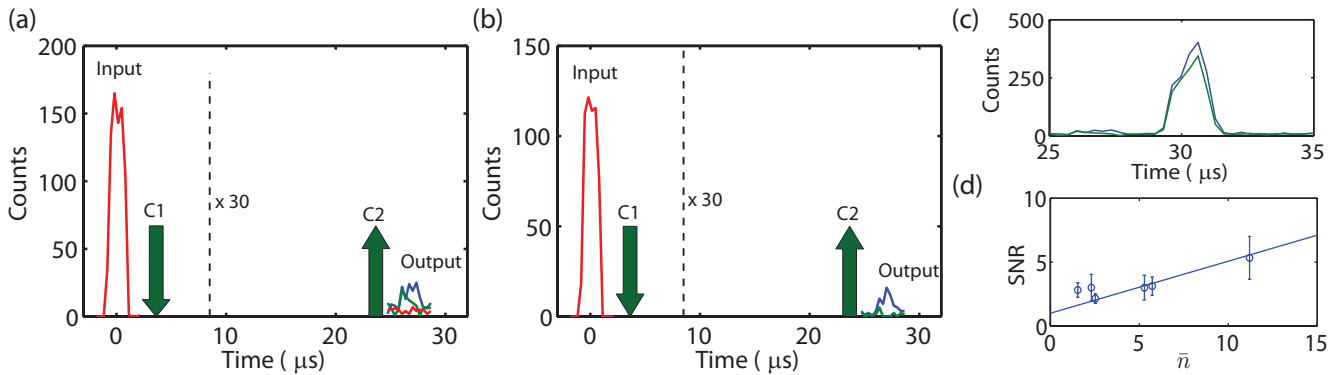}\\
\caption{Storage of a weak coherent pulse with (a) $\bar{n} = \nweak$ and (b) $\bar{n} = \nweakB$. The input mode recorded with no comb, the position of the control fields (C1 and C2), and finally a magnified (x 30) signal of the AFC spin-wave echo (blue curve), the associated noise without an input pulse (green curve) and the detector dark counts (red curve) are shown. Note that the total measurement time differs between the data sets in (a) and (b). (c) The same echo data of (a) with the temporally separated off-resonant echo (OREO). (d) SNR for different $\bar{n}$. Shown is also a fitted linear slope fixed to 1 at $\bar{n}=0$ by definition.}
\label{fig:3photon}
\end{figure*}

AFC spin-wave storage for weak coherent pulses with average photon numbers between  $\bar{n} = \nweak$ and $\bar{n} = \nweakB$ are shown in Fig. \ref{fig:3photon}. The input pulse is 2 $\mu$s long, the memory parameters are $1/\Delta$=6 $\mu$s and $T_S$=21 $\mu$s, leading to a total storage time of 27 $\mu s$. The duration and shape of the control pulses were optimized for the highest signal-to-noise ratio (SNR), see discussion below.
These measurements are performed, as all of the measurements shown in this letter, without the cryostat switched on to reduce the effect of vibrations on the comb structure \cite{Timoney2012}.

There are two principal mechanisms which are responsible for the noise created by the bright control pulses. One is scattering of the laser light itself from optical surfaces. Another is emission from the atoms which have been excited by the pulses, this includes incoherent fluorescence, coherent free-induction-decay (FID) type emission and an unexpected off resonantly excited echo.

Spatial separation of the input and control modes is used to shield the single photon counting detector from scattered light, but this did not lead to sufficient suppression. A double-pass AOM  (shown in Fig. \ref{fig:expsetup}b) is used as a detector gate in time, exploiting the temporal separation between the control fields and the emitted spin-wave echo, providing a suppression of roughly $10^6$. This proved sufficient to prevent detector blinding or significant afterpulsing.

The emission noise is, however, also present in the temporal mode of the output mode. A diffraction grating and a Fabry Perot (FP) cavity are used to spectrally filter this noise. The FP cavity is necessary in particular to reduce noise originating from FID. Sharp spectral features about the control transition, by products of the spectral tailoring required to prepare the AFC, cause the FID. We could reduce this noise by altering our preparation sequence, to increase the transparency window around the control field transition. The frequency of this noise is close to that of the control field, a fact which we observe by changing the frequency at which we lock our FP cavity.

In addition to the fluorescence and FID noise, we also observed an unexpected noise source at the input frequency. We believe its presence to be due to off-resonant excitation from the control fields, we thus call this an off-resonant echo (OREO) (see Fig. \ref{fig:3photon}c). The OREO is observed a time $1/\Delta$ after C2, supporting this hypothesis. We also observe a strengthening of the echo if C1 is present. We explain this by supposing that the off-resonant excitation of C1 is combined with transfer to the spin state. C2 then reads out the excitation in the same manner which it does the single-photon-level input pulse. The observed $1/\Delta$ dependence of the OREO, also fits to this explanation. Although the OREO is considerably larger than the AFC spin-wave echo which we are seeking to retrieve, the two echoes occur in temporally separated modes (see Figs \ref{fig:3photon}a and c). We could reduce the impact of temporal mode leakage by carefully tuning the shape of the control fields, which is consistent with an off-resonant excitation mechanism. Note that since the FID and the OREO are coherent processes, the corresponding emission should only be strong in the control mode. Scattering inside the crystal does however introduce significant cross talk between the spatial modes.

The temporal shape of the remaining noise we observe in Fig. \ref{fig:3photon}a is indicative of FID noise. This gives us reason to believe that a more efficient filtering system would permit us to increase the power in the control fields thus increasing their efficiency. The remaining noise, in this particular measurement, amounts to $(5.1\pm 1.3)\cdot10^{-3}$ photons per mode emitted at the crystal. The SNR up to $\bar{n} =\nweakB$ is shown in Fig. \ref{fig:3photon}d. These measurements were taken on a range of different days for the same experimental parameters. The SNR follows a linear dependence within the experimental errors, see fitted linear slope in Fig. \ref{fig:3photon}d. Measurements carried out for higher average photon numbers (not shown) confirmed this behaviour.

The final memory efficiency in the photon counting experiment was significantly lower than for the bright pulse storage. The optimization of the duration and shape of the control pulses led to a lower transfer efficiency. Furthermore, a photon counting experiment requires time averaging, for example, the measurement for $\bar{n} =\nweak$ was taken over the course of three hours. This challenges the stability of the experiment, in particular, laser fluctuations create reduced quality combs, which negatively affect the AFC echo efficiency. Averaging over all the measurements shown in Fig. \ref{fig:3photon}d, we obtain a global memory efficiency of \mbox{\etas} and an unconditional noise floor of \pnoise.

\begin{figure}
{\includegraphics[width = 0.48\textwidth]{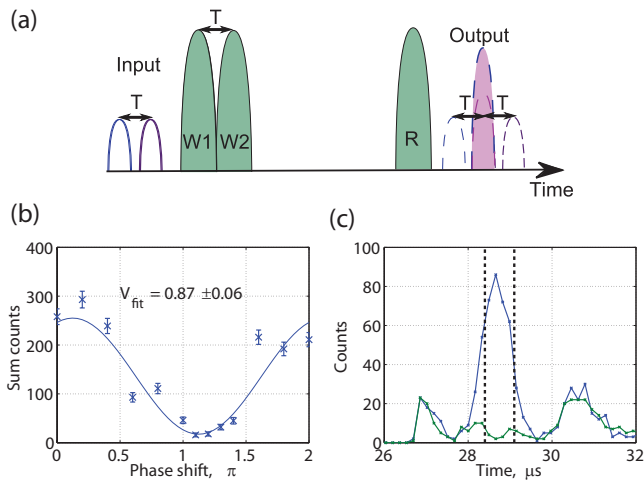}}
\caption{
(a) The method used to measure the coherence of the AFC spin-wave echoes.
The single write operation of Fig. \ref{fig:3photon}a,b (C1) is replaced by a double write operation(W1, W2). If the temporal separation (T) of the input mode  is equal to that of the double write operation , the first echo of the second write operation and second echo of the first operation interfere.
(b) The visibility curve for two pulses with $\bar{n} = 176 \pm 8$ . (c) The signal of a constructive and destructive case. The thick dashed lines show the temporal window which was used to obtain the interference curve. The detector gate has cut some of the first echo and the OREO is not shown in this temporal slice. }
\label{fig:coherence}
\end{figure}

Finally we show the coherence of the AFC spin-wave echo. To do this we store a time-bin pulse in the memory where we vary the phase of one of the time bins. We then self interfere the time-bin pulse using a temporal beamsplitter and examine the interference curve. The visibility of the curve gives a measure of coherence preservation in the memory. For the measurement shown in this letter, the temporal beamsplitter comes in the form of the control pulses. The scheme is pictorially shown in Fig. \ref{fig:coherence}a.

To store and analyse the time-bin pulse, we need clean temporal separation between the retrieved pulses and enough time to see the triple pulse structure shown in Fig. \ref{fig:coherence}c after the final control field. To do this we extend the AFC time from 6 to 8 $\mu$s, and reduce the pulse width of the input pulses and the entire pulse length of the control pulses. These measures further reduce the efficiency with which we can store in the memory to $\eta_{s} $ = $(6.3 \pm 0.1)\cdot 10^{-4}$ for each mode, including the reduction in storage efficiency due to the double write operation. $T_S$ was set to 21 $\mu$s in this experiment, yielding a total memory time of about 29 $\mu$s. A visibility curve for $\bar{n} =176\pm 8$ is shown in figure \ref{fig:coherence}b, where we measure V = 0.87 $\pm 0.06$. We suspect that laser phase noise contributes negatively to our visibility curve. A simple calculation shows that frequency noise with $\sigma_f = $ 25 kHz reduces the baseline to V = 0.95. Together with the noise level this accounts for the visibility we measure. For $\bar{n} =51\pm3$ we observe a further drop in visibility to V = 0.71 $\pm 0.1$. This is due to the increasingly important role of noise in determining the minimum of the visibility curve. We note that with higher storage efficiency, it should be possible to obtain high visibilities for lower photon numbers.

The unconditional noise floor achieved in our experiment should in principle allow us to store a single-photon-level optical pulse with high SNR. The limited SNR obtained at a few photons is entirely given by the low overall memory efficiency. Future experiments should therefore aim at increasing the efficiency, while we consider the filtering to be sufficient for quantum applications. The memory inefficiency is due to two major factors; 1) insufficient optical depth and 2) insufficient control field transfer efficiency. To increase the optical depth we will implement an impedance-matched cavity around the memory, as proposed in \cite{Afzelius2010a, Moiseev2010a}. Indeed recent results using such an impedance-matched cavity have shown an optical AFC efficiency of 58$\%$ using a crystal with optical depth comparable to ours if in a single-pass configuration \cite{Sabooni2013}. 2) The control field transfer efficiency can most easily be improved by using longer adiabatic transfer pulses \cite{Minar2010}. Such long temporal windows can be created by increasing the AFC echo time ($1/\Delta$). Using the narrowest measured optical homogeneous linewidth measured with europium of 122 Hz \cite{Equall1994}, $1/\Delta$ times of around 1 ms are in principle possible. Furthermore increasing $1/\Delta$ will also allow us to exploit the multimode capability of the AFC scheme. Currently, however, our laser linewidth represents serious technical obstacle in increasing $1/\Delta$ towards this limit, beyond the shown 6-8 $\mu$s. We also note that employing long adiabatic control fields should reduce off resonant excitation, further decreasing the remaining noise.

To conclude, we have demonstrated the first optical storage as a spin-wave in a solid-state memory, in the regime of a few photons per input pulse. This was made possible by a strategy of extensive filtering and by carefully shaping the temporal envelope of the strong control pulses. The final unconditional noise floor of \pnoise $~$is low enough to allow for quantum schemes using spin-wave storage and manipulation, such as the generation of quantum-correlated spin-wave and photonic excitations using variant of the DLCZ \cite{Duan2001} approach adapted to the solid-state \cite{Ledingham2010, Beavan2012, Sekatski2011}. These schemes will, in turn, allow for generation of entanglement between light and matter and entanglement of solid state remote quantum memories, a basic building block for quantum repeaters.

We would like to thank C. Barreiro for technical assistance, and N. Sangouard and P. Goldner for useful discussions. We gratefully acknowledge R. Cone and R.M. MacFarlane for lending us the isoptically pure $^{151}$Eu crystal. This work was financially supported by the Swiss National Centres of Competence in Research (NCCR) project Quantum Science Technology (QSIT) and by the European projects QuRep (FET Open STREP), CIPRIS (FP7 Marie Curie Actions) and Q-Essence (FET Proactive Integrated Project).

\end{document}